\documentclass[aps,prb,twocolumn,amsmath,amssymb,superscriptaddress,nobibnotes,showpacs,floatfix]{revtex4-1}
\usepackage{graphicx,SIunits,array,multirow,amsmath,amsfonts,tabularx}
\usepackage{epstopdf}
\usepackage{dsfont,color}

\newcolumntype{C}[1]{>{\centering\let\newline\\\arraybackslash\hspace{0pt}}m{#1}}

\newcolumntype{Y}{>{\centering\arraybackslash}X}

\begin{document}
\preprint{}

\title[]{Magnetic properties of bilayer Sr$_3$Ir$_2$O$_7$: role of epitaxial strain and oxygen vacancies}

\author{Bongjae Kim}
\affiliation{University of Vienna, Faculty of Physics and Center for Computational Materials Science, Vienna, Austria}
\author{Peitao Liu}
\affiliation{University of Vienna, Faculty of Physics and Center for Computational Materials Science, Vienna, Austria}
\affiliation{Institute of Metal Research, Chinese Academy of Sciences, Shenyang 110016, China}
\author{Cesare Franchini}
\affiliation{University of Vienna, Faculty of Physics and Center for Computational Materials Science, Vienna, Austria}

\date[Dated: ]{\today}

\begin{abstract}
Using {\it ab initio} methods, we investigate the modification of the magnetic properties of the $m=2$ member of the strontium iridates
Ruddlesden-Popper series Sr$_{m+1}$Ir$_{m}$O$_{3m+1}$, bilayer Sr$_3$Ir$_2$O$_7$,  induced by epitaxial strain and oxygen vacancies. Unlike the single layer compound Sr$_2$IrO$_4$, which exhibits a robust in-plane magnetic order, the energy difference between in-plane and out-of-plane magnetic orderings in Sr$_3$Ir$_2$O$_7$ is much smaller and it is expected that small external perturbations could induce magnetic transitions.
Our results indicate that epitaxial strain yields a spin-flop transition, that is driven by the crossover between the intralayer $J_1$ and interlayer $J_2$ magnetic exchange interactions upon compressive strain. While $J_1$ is essentially insensitive to strain effects, the strength of $J_2$ changes by one order of magnitude for tensile strains $\geq$ 3~\%. In addition, our study clarifies that the unusual in-plane magnetic response observed in Sr$_3$Ir$_2$O$_7$ upon the application of an external magnetic field originates from the canting of the local magnetic moments due to oxygen vacancies, which locally destroy the octahedral networks - thereby allowing for noncollinear spin configurations.
\end{abstract}


\maketitle

\section{Introduction}

\begin{figure}[t]
\begin{center}
\includegraphics[width=85mm]{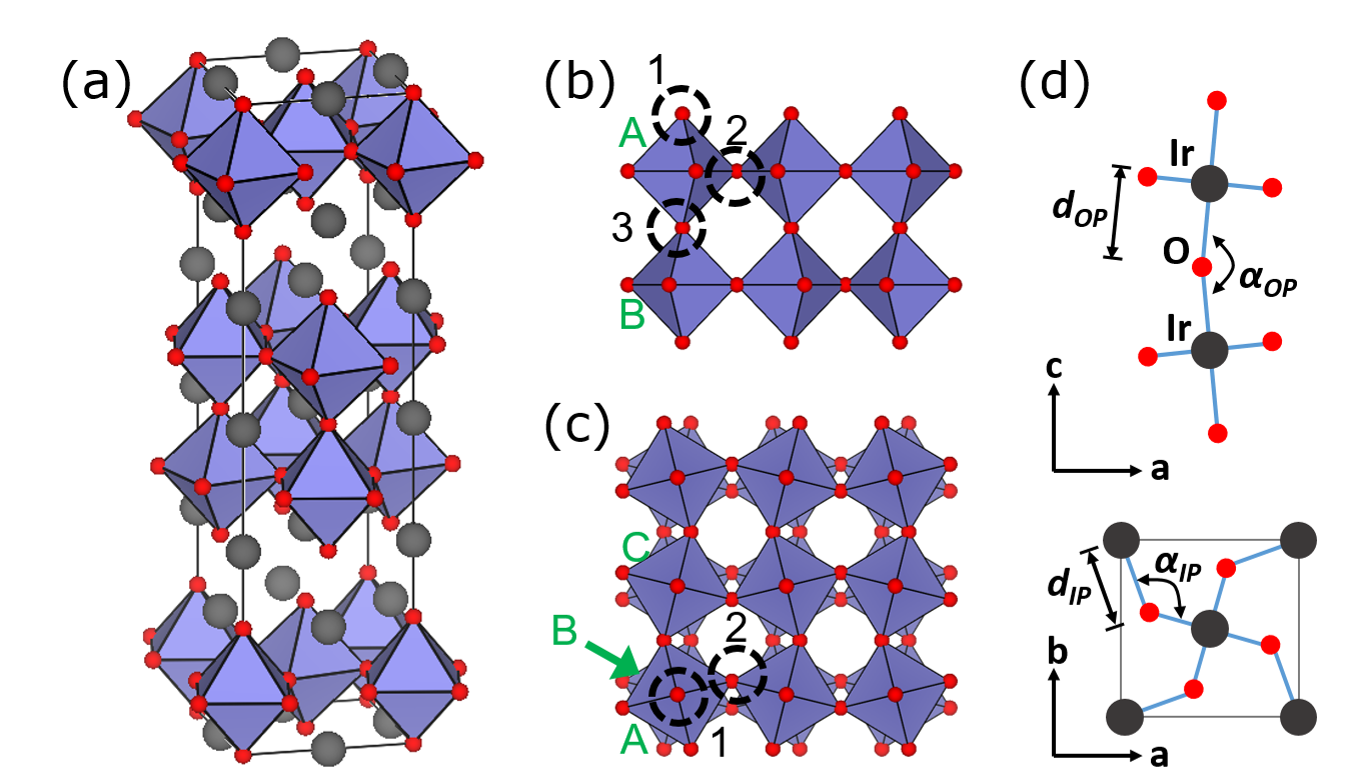}
\caption{(Color online) Structure and vacancy positions of Sr$_3$Ir$_2$O$_7$.
(a) Crystal structure of Sr$_3$Ir$_2$O$_7$. IrO$_6$ octahedra are denoted with polyhedra, and each O and Sr ion with small (red) and large (gray) spheres, respectively.
(b) Side view and (c) top view of RP perovskite bilayer with $\sqrt{2} \times \sqrt{2}$ 2D unit cell. Each oxygen vacancy (Ov) position is marked with a dashed circle. (d) Schematic description of the Ir-O-Ir bond angle and Ir-O bond length of both apical ($\alpha_{OP}$, $d_{OP}$) and planar ($\alpha_{IP}$, $d_{IP}$) directions.
}
\label{structure}
\end{center}
\end{figure}

 The discovery of the relativistic Mott insulating phase in 5$d$ oxide Sr$_2$IrO$_4$, originating from the delicate balance of electronic correlation ($U$), bandwidth ($W$), and spin-orbit coupling (SOC)~\cite{BJKim2008,BJKim2009}, has attracted great interest from the condensed matter physics community. Sr$_2$IrO$_4$ is the first ($m=1$) member of the Ruddlesden-Popper (RP) series Sr$_{m+1}$Ir$_{m}$O$_{3m+1}$, $m=1,2,\cdots \infty$. This RP family has been the subject of many studies focused on clarifying the role of dimensionality on the electronic and magnetic properties~\cite{Moon2008,JWKim2012,Carter2013,Zhang2013}. The RP phase of strontium iridates Sr$_{m+1}$Ir$_{m}$O$_{3m+1}$ is composed of $m$ two-dimensional (2D) layers of IrO$_6$ corner-sharing octahedra joined along the perovskite stacking direction and separated by rock-salt SrO layers
 (See Fig.~\ref{structure}(a) for $m=2$ case). As the electronic structure and magnetic properties of the systems are mostly determined by the Ir-$d$ states and their hybridization with O-$p$ states, the IrO$_2$ layers within the perovskite blocks play a crucial role in determining the material properties of the system: as $m$ increases, the number of the interlayer hopping paths between IrO$_2$ layers increases and this enhances the degree of electronic and magnetic itinerancy of the system. Many studies have discussed the crossover between 2D to three-dimensional (3D) behaviors~\cite{Moon2008,Carter2013,Zhang2013}. Recently, cuprate-like electronic structures in the 2D limit
 ($m=1$, Sr$_2$IrO$_4$)~\cite{YKKim2014,Yan2015,YKKim2015} and topological characteristics in the 3D limit ($m=\infty$, SrIrO$_3$)~\cite{Carter2012,Zeb2012,JLiu2016} were reported, that have introduced even more fuel on current researches on RP-structured strontium iridates.

 Among the members of the RP series, the $m=2$ bilayer system Sr$_3$Ir$_2$O$_7$, is located in a unique intermediate position between the 2D and 3D limit
and is characterized by peculiar electronic and magnetic properties which are different from those
 of the other members of the RP series. Single layer Sr$_2$IrO$_4$ has a moderate gap and a canted in-plane (IP) magnetic order, whereas
Sr$_3$Ir$_2$O$_7$ has a narrower charge gap~\cite{Okada2013,Park2014},
 and shows collinear out-of-plane (OP) magnetic structure~\cite{JWKim2012,Boseggia2012_2} with an unusually large magnon gap~\cite{JKim2012}.
 The magnetic ordering of the IP and OP configurations are schematically given in Fig.~\ref{E_U} (a).
 The IP-to-OP magnetic transition going from  Sr$_2$IrO$_4$ to Sr$_3$Ir$_2$O$_7$
 is claimed to be due to the modification of the interlayer exchange interaction between IrO$_2$ layers~\cite{JWKim2012}. This implies that by controlling the strength of the interlayer interaction, one could tune the magnetic structure of the system to
 give spin-flop transition without dimensional changes. One way to modify the magnetic interactions is to tune the distance between the bilayers and
the connectivity of the IrO$_6$ octahedra by epitaxial strain; this can be achieved experimentally
by making use of different substrates~\cite{Rondinelli2011}. Similarly to previous studies on strain effects in the $m=1$ and $m=\infty$ series~\cite{Nichols2013,Serrao2013,Lupascu2014,BKim2016_1,JLiu2013,Biswas2014,Zhang2015,JLiu2016}, epitaxial strain study in the $m=2$ system is expected to give new insights into physics of RP iridates.

 Another peculiar and still unresolved issue in Sr$_3$Ir$_2$O$_7$ is the observation of an unusual IP magnetic response at low temperature upon the application of an external magnetic field which is apparently in contrast with the well established OP magnetic order determined from X-ray diffraction spectroscopy~\cite{Cao2002,JWKim2012}. Possible non-collinear magnetic order or canting of the moments were proposed to explain this intriguing
 IP magnetic behavior~\cite{Dhital2012,Boseggia2012}. Recently, based on a refined analysis of the Sr$_3$Ir$_2$O$_7$ crystal structure, it was proposed that the tilting of the IrO$_6$ octahedra could be responsible for the observed IP signal~\cite{Hogan2016}. According to the reported growth phase diagram~\cite{Nishio2016}, obtaining strontium iridates with perfect stoichiometry is very difficult and oxygen vacancies (Ovs) are easily formed with high tendency of intermixing in the $m=1$ (IP) and $m=2$ (OP) phases~\cite{Sung2016}, which suggest the idea that external impurities or vacancies could be responsible for the IP magnetic behavior~\cite{Fujiyama2012}.

 In this study, by using {\it ab initio} approaches based on relativistic density functional theory (DFT) plus an on-site Hubbard $U$, we address the above mentioned issues on the magnetic structure of Sr$_3$Ir$_2$O$_7$: tunability of the spin-flop transition and origin of the IP magnetic response.
We find that the system can be subjected to spin-flop transitions driven by the interlayer exchange interaction between IrO$_2$ layers. Also, the role of the Ovs is investigated in relation with the reported IP magnetic behavior and we propose that the Ovs can in fact induce the canting-off of the collinear OP Ir magnetic moment.

\section{Calculation details}

 We have performed {\it ab initio} electronic structure calculation employing the projector augmented wave method implemented in the Vienna {\it ab initio} simulation package (VASP)~\cite{Kresse1993,Kresse1996}. We used the generalized gradient approximation (GGA) of Perdew-Becke-Ernzerhof (PBE) for the exchange-correlation functional~\cite{Perdew1996}. The electronic correlation of the Ir-$d$ orbitals is treated within the GGA+$U$ method with full consideration of SOC effect~\cite{Dudarev1998}. The bulk system, as described in Fig.~\ref{structure} (a), is fully relaxed and resulted in a small ($<3\%$) overestimation of the volume with respect to the experimental one. For the strained system, the in-plane lattice parameters are fixed to the corresponding substrate and full relaxation of the out-of-plane lattice parameters and the internal atomic positions is performed at standard convergence criteria. To quantify the $U$ parameter of the system, we have performed constrained random phase approximation (cRPA) calculations using a Ir-$t_{2g}$ basis set projected onto Wannier orbitals~\cite{Aryasetiawan2006,Marzari1997,Mostofi2008,Franchini2012}. We obtained an effective $U$ value of 1.6 eV. The calculated on-site Coulomb ($U_{ij}$) and exchange ($J_{ij}$) interaction parameters within Ir-$t_{2g}$ orbitals are shown in Table~\ref{crpa}, where $i$ and $j$ are the orbital index of the $t_{2g}$ manifolds. The effective $U$ value used in the DFT+$U$ runs is obtained from the averaged difference of Coulomb and exchange interactions ($\bar{U}$-$\bar{J}$). Unless specified, we have employed our obtained $U$ of 1.6 eV for all calculations.
 For the strain effect, we employed unit cell described in Fig.~\ref{structure}(a), which contains 4 formula unit (f.u.) of Sr$_3$Ir$_2$O$_7$.
To describe the Ovs, we have considered a $\sqrt{2} \times \sqrt{2}$ supercell (SC) containing 8 formula units.
This correspond to a Ov concentration of about 1.8~\%, very close to the intrinsic Ov concentration for nearly
stoichiometric strontium iridates extracted from experiment, 1.0~\%.
We have inspected the three inequivalent positions for Ovs shown in Fig.~\ref{structure} (b) and (c).
A Monkhorst-Pack $k$-points 6$\times$6$\times$2 mesh was used for 4 f.u. cell, which is reduced to 4$\times$4$\times$2 for the SC case. For the description of the final density of states and magnetic moments, we have adopted a denser 6$\times$6$\times$3 mesh.

\begin{table}[b]
\centering
\caption[]{
Calculated on-site Coulomb ($U_{ij}$) and exchange ($J_{ij}$) interaction parameters within Ir-$t_{2g}$ orbitals based on the cRPA calculations. $i$ and $j$ are the orbital index from $t_{2g}$ manifolds.
}
\begin{tabular}{C{0.9cm}|C{1.0cm}C{1.0cm}C{1.0cm}|C{0.9cm}|C{1.0cm}C{1.0cm}C{1.0cm}}
\hline\hline
  $U_{ij}$  & $xy$  & $xz$ & $yz$ &     $J_{ij}$ & $xy$  & $xz$ & $yz$ \\
\hline
  $xy$        & 2.30  & 1.71 & 1.53 &     $xy$       & -     & 0.22 & 0.21 \\
  $xz$        & 1.71  & 2.29 & 1.53 &     $xz$       & 0.22  & -    & 0.21 \\
  $yz$        & 1.53  & 1.53 & 1.92 &     $yz$       & 0.21  & 0.21 & -    \\
\hline\hline
\end{tabular}
\label{crpa}
\end{table}

\section{Results and discussions}

\subsection{Bulk pristine Sr$_3$Ir$_2$O$_7$ case}

\begin{figure}[t]
\begin{center}
\includegraphics[width=85mm]{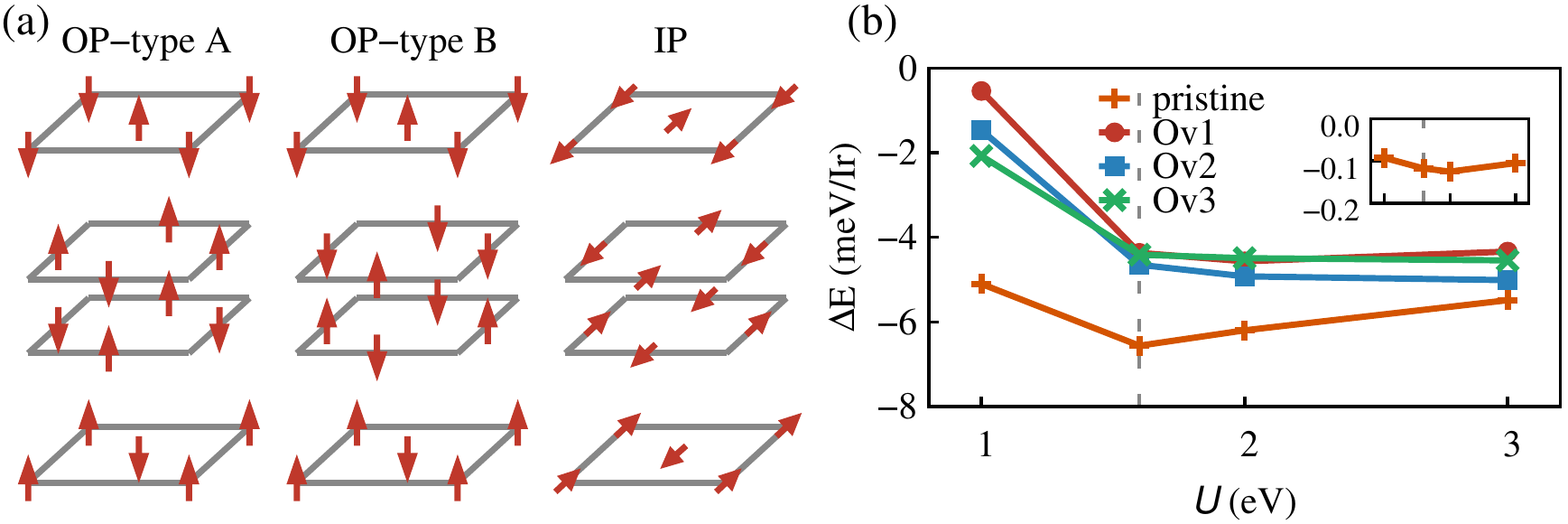}
\caption{(Color online)
(a) Schematic magnetic structures of the two OP orderings, type A and B, proposed in literature~\cite{Boseggia2012_2}, and the IP one. According to our calculations the B-type order corresponds to the ground state phase and it is only 0.1 meV/atom more stable than the A-type one.
(b) The energy difference between collinear OP magnetic order and noncollinear IP magnetic order ($\Delta E = E_{OP}-E_{IP}$) for both pristine sample and sample with oxygen vacancy. The overall decrease of the $\Delta E$ for $U$=1 eV
is due to decrease of the Ir magnetic moment. Inset: zoom of the energy difference between OP-A and OP-B.
}
\label{E_U}
\end{center}
\end{figure}

 We first examine the electronic structure and magnetic properties of bulk pristine Sr$_3$Ir$_2$O$_7$. We found a band gap of 0.29 eV, slightly larger than the measured values reported in literature~\cite{Okada2013,Wang2013}. In line with previous studies~\cite{Zhang2013,King2013}, we found that the system deviates from the $J_{eff}=1/2$ description characteristic of Sr$_2$IrO$_4$ ($m=1$), which is evident from the computed $\mu_L$/$\mu_S$ ratio ($\mu_L$ and $\mu_S$ being the orbital and spin moment) of 1.2, a value which is much lower than the ideal value of 2 expected for a $J_{eff}=1/2$ state. As shown in Fig.~\ref{E_U}, our data confirm that the OP magnetic structure is more stable than the IP one by a few meV/atom for reasonable $U$ values ranging from 1 to 3 eV, in good agreement with former experimental reports~\cite{Boseggia2012_2,JWKim2012}.
 For small $U$, it can be seen that the energy difference $\Delta{E}$ between OP and IP orderings is smaller. This is probably due to the fact that
for small $U$ the system is pushed closer to the boundary of an insulator-to-metal transition with negligible magnetic moments. Noteworthy, being the energy difference between OP and IP magnetic arrangements only few meV per Ir atom, substrate engineering can be exploited to induce spin-flop transition as we will discuss later on.

 As described in Fig.~\ref{E_U} (a), there are two alternative OP magnetic arrangements that preserve antiferromagnetic configurations within the IrO$_2$ bilayers, that differ only in the inter-bilayer magnetic structure. These are denoted as OP-type A and OP-type B.
Experimental ambiguities on the specific type of OP ordering remain:
Kim {\it et. al.} reported the OP-type B (see Fig.~\ref{E_U} (a)) to be the correct magnetic ground state~\cite{JWKim2012}, while Boseggia {\it et. al.} reported that both OP-A and -B magnetic setups appear to be consistent with X-ray resonant scattering experiments~\cite{Boseggia2012_2}. Our data confirm that type B is energetically more stable than type A by only $\simeq$ 0.1 meV per Ir atom, which is robust upon tested $U$ values between 1 to 3 eV, and as we will show later, also upon epitaxial strain and different types of Ovs. This energy difference is much smaller than the one between IP and OP, indicating that the bilayer-bilayer interaction is weak and the system still possesses a certain 2D magnetic nature. Moreover, the fact that the two OP magnetic structures are almost degenerate in energy supports the idea of possible multi-domain behaviors in Sr$_3$Ir$_2$O$_7$ as suggested by Boseggia {\it et. al.}~\cite{Boseggia2012_2}.

\subsection{Epitaxial strain effect}

 We now address the effect of epitaxial strain. For the $m=1$ system, 2D Heisenberg-type model approaches suggested a possible IP to OP spin-flop transition  associated with a change of the crystal field splitting parameters~\cite{Jackeli2009}. However, a recent report based on magnetically noncollinear DFT reported that the  tetragonal distortion required to flop the spins is rather large~\cite{PLiu2015}, which might indicate that
a local picture is not sufficient for the exact estimation of the crystal field of the system~\cite{Bogdanov2015}.
This was also confirmed by recent calculations concluding that the IP order is robust for large ranges of epitaxial strain~\cite{BKim2016_1}.

 The situation is different for the $m=2$ system. The additional IrO$_2$ layer in Sr$_3$Ir$_2$O$_7$ induces a dimensionality-driven spin-flop transition from canted IP to collinear OP order, which is explained by the emergence of Heisenberg-type interlayer interactions~\cite{JWKim2012}. In the case of Sr$_3$Ir$_2$O$_7$, the energy difference between the IP and OP magnetic orderings is only a few meV/Ir, which is in stark contrast
 to Sr$_2$IrO$_4$ that exhibits a robust IP magnetic order. This suggests that small external perturbations, such as epitaxial strain or electron/hole doping, could induce spin-flop transitions by changing the strength of the effective magnetic interactions.

\begin{figure}[t]
\begin{center}
\includegraphics[width=85mm]{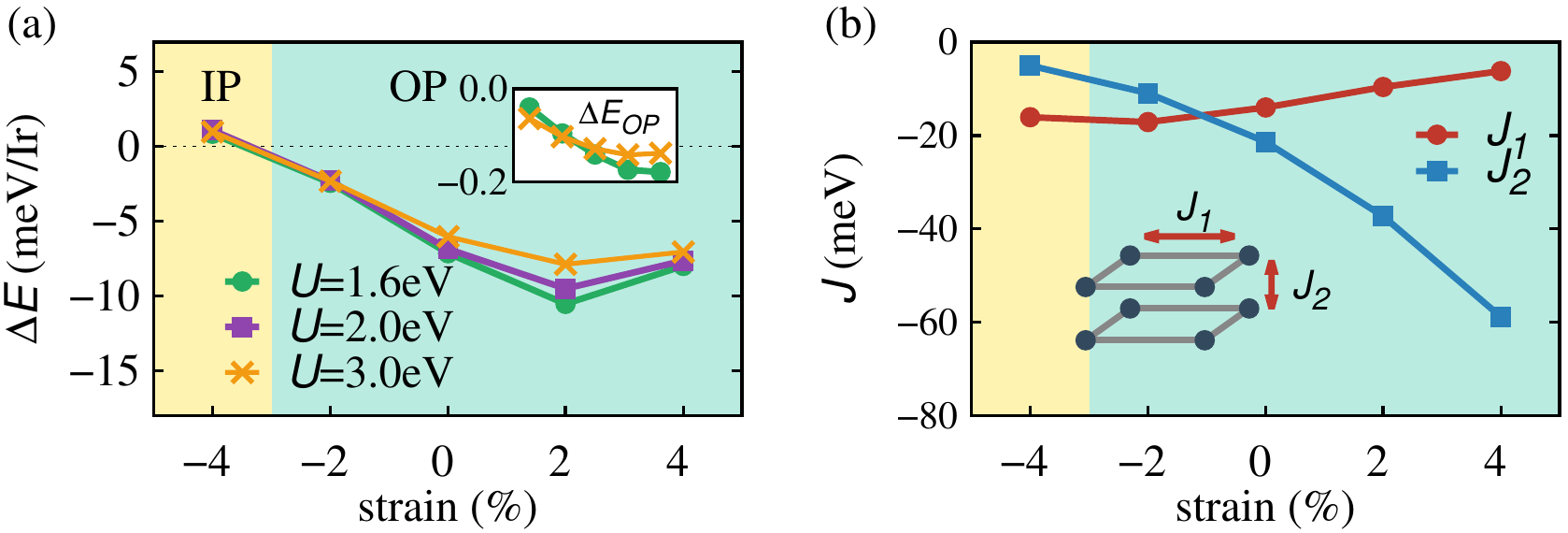}
\caption{(Color Online)
(a) Energy difference $\Delta E$ between the IP and OP magnetic orders as a function of the epitaxial strain. A switch between the IP and OP magnetic orders is found at around -3\% strain.
Inset: zoom of the  energy difference between OP-A and OP-B.
(b) Calculated $J$ parameters upon strain. (Inset) Schematic view of the bilayer lattice.  $J_1$ and $J_2$ are the planar and apical exchange interactions, respectively.
}
\label{E_J}
\end{center}
\end{figure}

 In our study, we have modelled the substrate-induced epitaxial strain by changing the in-plane lattice constant, a standard tuning tool to investigate
the effect of strain in computational experiments. Among many possible substrates, we have chosen SrTiO$_3$ (STO) as a reference substrate since the lattice mismatch between Sr$_3$Ir$_2$O$_7$ and STO is almost zero~\cite{BKim2016_2}. The IP lattice parameters of Sr$_3$Ir$_2$O$_7$ are fixed to those of the STO substrate (this represents our zero, i.e. 0\% strain), then both compressive
 (-2\%, and -4\%) and tensile strain (2\% and 4\%) cases were systematically investigated. Both OP and IP magnetic structures were inspected. The results are summarized in Fig.~\ref{E_J}. Without strain (0\%), corresponding to the STO substrate case, the OP order is favored over the IP one by 7 meV/Ir, which is very similar to bulk situation (Fig.~\ref{E_J} (a)). For tensile strain up to 4\%, the OP order is energetically favored, but for
 compressive strain  $\geq 3\%$, the IP phase becomes more stable eventually yielding a spin-flop transition from OP to IP (Fig.~\ref{E_J} (a)).

 The strain-driven spin-flop behavior found for this bilayer system ($m=2$) is very unique as there are no such examples in the whole RP series of strontium iridates. Note that for Sr$_2$IrO$_4$ the IP order is very stable and conventional perturbations such as realistic changing of the crystal field, substrate strain or hole/electron doping do not guide a change of the magnetic order. Only a selective substitutional doping at magnetic site could influence the type of magnetic ordering~\cite{PLiu2016_2,delaTorre2015,Lupascu2014,Lado2015,BKim2016_1,Calder2012}. Also, for SrIrO$_3$ (m=$\infty$), the nonmagnetic metal character of the system is preserved for a wide range of substrate strains~\cite{JLiu2016,BKim2016_1}.

\begin{table}[b]
\centering
\caption{Ir-O-Ir bond angle and Ir-O bond length for different strains. Both apical (OP) and planar (IP) bond angle and bond length are schematically described in Fig.\ref{structure} (d). Also, Ir-Ir distances and IP/OP lattice parameter ($a$ and $c$) based on the structure shown in Fig.\ref{structure} (a) are provided. Angles are in degree ($^{\circ}$) and length scales are in \AA}.
\begin{tabular}{C{1.0cm}|C{1.0cm}|C{0.9cm}|C{0.9cm}C{0.9cm}C{0.9cm}C{0.9cm}C{0.9cm}}
\hline\hline
                          &               &     bulk    &    -4\%  &    -2\%  &    0\%    &    2\%    &    4\%      \\
\hline
\multirow{3}{*}{planar} & $\alpha_{IP}$  &   152.3   &  147.9   &  150.4   &   152.9   &   155.2   &   157.1     \\
                          & $d_{IP}$     &   2.00    &  1.95    &  1.98    &   2.01    &   2.04    &   2.07      \\
                          & $d_{Ir-Ir}$  &   3.93    &  3.75    &  3.83    &   3.91    &   3.98    &   4.06      \\
\hline
\multirow{3}{*}{apical}   &$\alpha_{OP}$ &   180.0    &\multicolumn{5}{c}{180.0}       \\
                          & $d_{OP}$     &   2.05    &  2.11    & 2.07     & 2.04      & 2.01      & 1.98     \\
                          & $d_{Ir-Ir}$  &   4.13    &  4.22    &  4.15    &   4.08    &   4.02    &   3.96      \\
\hline
\multicolumn{2}{c|}{$a$}                &     5.56   &    5.30 &    5.41 &   5.52   &   5.63   &   5.74     \\
\multicolumn{2}{c|}{$c$}                &    21.21   &   21.57 &   21.25 &  20.93   &  20.63   &  20.34     \\
\hline\hline
\end{tabular}
\label{bond}
\end{table}

 To understand the origin of the strain-driven spin-flop transition, we have calculated the intralayer and interlayer exchange interaction parameters ($J_1$ and $J_2$: see the inset of Fig.~\ref{E_J} (b)), which are extracted from DFT total energies of various magnetic orders and different strains assuming Heisenberg-type interactions between Ir sites. In order to stabilize all different spin configurations required to compute the exchange interactions and to guarantee an accurate degree of convergence we have used $U$=3~eV. This is not expected to change significantly the value of the interaction parameters with respect to those obtained using the cRPA value of $U$, 1.6 eV, since the strain-dependent total energy curves are not largely dependent on $U$ (see Fig.~\ref{E_J} (a)). As expected, the exchange interactions between neighboring Ir sites are found to be antiferromagnetic (negative values for both exchange interactions, see Fig.~\ref{E_J} (b)). The strength of the magnetic interactions are several tens of meV/Ir, in agreement with previous reports on various iridates obtained by different approaches~\cite{JKim2012,Katukuri2012,BHKim2012,Katukuri2014}.

 Starting from the most compressive case, as tensile strain is applied, there is a decrease (increase) of $J_1$ ($J_2$) due to the increase of the planar (apical) Ir-O distance; this can be interpreted in a tight-binding picture as a reduction (enhancement) of the intersite hopping amplitude $t$ between Ir sites, according to  the relation $J_i \sim t^2/U$.
Interestingly enough, as shown in Fig.~\ref{E_J} (c), the change in $J_2$ is very large compared to the $J_1$. The global decrease of $J_1$ from -4\% to 4\% is only about $\sim$ 10 meV/Ir while the increase of the $J_2$ is more than 50 meV/Ir. The origin of this different response of $J_1$ and $J_2$ upon strain lies in the geometrical atomic connectivity of the system: for the planar interaction $J_1$, as the tensile strain increases, \emph{both} the Ir-O bond lengths and Ir-O-Ir angles increase and the reduction of the hopping due to change in bond length is compensated by the enhanced hopping due to the rectification of the Ir-O-Ir angle (From 147.9$^{\circ}$ to 157.1$^{\circ}$, see Table~\ref{bond}). For the interlayer interaction $J_2$, however, the Ir-O-Ir angle  remains 180$^{\circ}$ regardless of the strain, and the reduction of apical Ir-O length enhances the exchange interaction more efficiently.
 Considering the predominantly quasi-2D character of the system, where the electrons are more spatially confined, the interlayer interaction can in turn be  highly dependent on the variation of the Ir-O distance.

 Previous theoretical results based on tight-binding calculations~\cite{Carter2013}, microscopic model approaches~\cite{JWKim2012}, and recent angle-resolved photoemission spectroscopy~\cite{Moreschini2014} have suggested the importance of the interlayer coupling in determining the magnetic structure of Sr$_3$Ir$_2$O$_7$, which is explicitly shown in our calculations. Note that the tiny energy difference between the type A and B orderings is almost insensitive to strain, implying that the role of the long-range neighbor exchange interactions should be minimal and our analysis based on the two short-range and strongest interactions, $J_1$ and $J_2$, is valid.

\subsection{Role of oxygen vacancies}

\begin{figure}[b]
\begin{center}
\includegraphics[width=85mm]{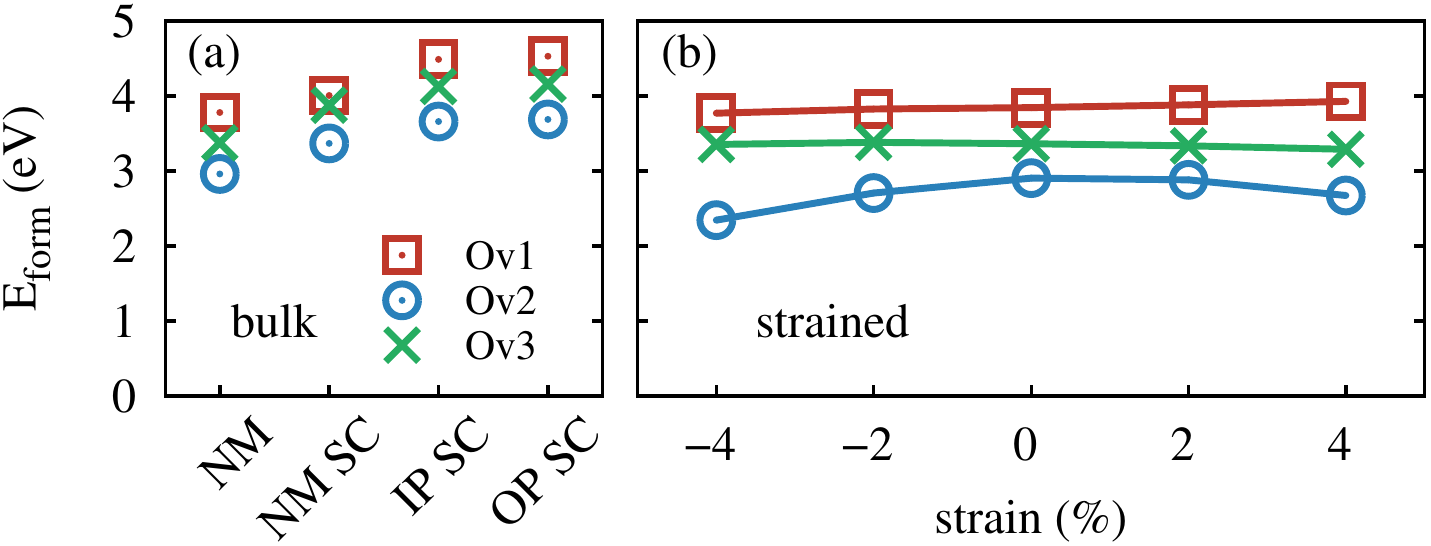}
\caption{(Color online)
(a) Formation energy of an oxygen vacancy ($\rm {E_{form}}$). NM, NM SC, IP SC, and OP SC , respectively, refer to nonmagnetic unit cell, nonmagnetic supercell, supercell with IP magnetic order, and supercell with OP magnetic order (OP SC). (b) Strain dependency of $\rm {E_{form}}$.
}
\label{FE}
\end{center}
\end{figure}

 Both experimental measurements and theoretical calculations clearly show that the ground state magnetic structure of Sr$_3$Ir$_2$O$_7$ is OP-type~\cite{Carter2013,JWKim2012,Boseggia2012_2}. However, the origin of the experimentally observed unusual magnetic response along the IP direction~\cite{Cao2002} remains a debated issue that has not been resolved so far. Various possible explanations have been suggested involving external defects~\cite{Fujiyama2012}, canting-off of the local moments~\cite{JWKim2012,Boseggia2012,Hogan2016}, multiple domains~\cite{Dhital2012},
 and noncollinear magnetic order~\cite{Dhital2012} but none of these came up with a firm solution. Recent studies on the synthesis of strontium iridates has shed some light on the issue. It was reported that the magnetic properties of the systems are highly dependent on the complicated growth conditions, and Ovs were found to be responsible for divergent reports from different experiments~\cite{Sung2016,Nishio2016}. Moreover, the narrower stability window in the growth phase diagram of Sr$_3$Ir$_2$O$_7$ makes the system more susceptible to defects~\cite{Nishio2016}, and phase intermixing with Sr$_2$IrO$_4$~\cite{Sung2016} can occur. Therefore, Ovs can be thought to be primal candidates responsible for the unusual IP magnetic response in Sr$_3$Ir$_2$O$_7$. In order to elucidate this issue we have studied the effect of Ovs considering  different Ovs positions, as depicted in  Fig.~\ref{structure}(b) and (c): we have inspected the outer apical site (Ov1), the inner IP site (Ov2), and the inner apical site (Ov3) within the bilayer block. The most favorable Ov site is identified from the formation energy, calculated by using the following formula:
\begin{equation}
 E_{form} = E_{tot,O_v}-E_{tot}+\mu_{O},
\label{form}
\end{equation}
 where $E_{tot,O_v}$, $E_{tot}$, and $\mu_{O}$ denotes the total energy of the system with and without Ov, and the chemical potential of oxygen, respectively. Here, the chemical potential of oxygen is obtained from the energy of oxygen in an isolated O$_2$ molecule ($\mu_{O}=\frac{1}{2}\mu_{O_2}$).
 This approach is known to give good agreement with experiments for perovskite oxides~\cite{Aschauer2013}. The calculations were done for the original unit cell (Fig.~\ref{structure}(a)) with nonmagnetic (NM) configurations and for the SC (Fig.~\ref{structure}(b) and (c)) with NM, IP and OP magnetic configurations. The results, summarized in Fig.\ref{FE}(a), show that the Ov2 is the most favorable case regardless of the cell size and the magnetic order. Also, both compressive and tensile strain were found not to alter the
 relative stability of Ovs, confirming that Ov2 represents the optimal site for oxygen vacancy formation (Fig.~\ref{FE} (b)).

\begin{figure}[t]
\begin{center}
\includegraphics[width=85mm]{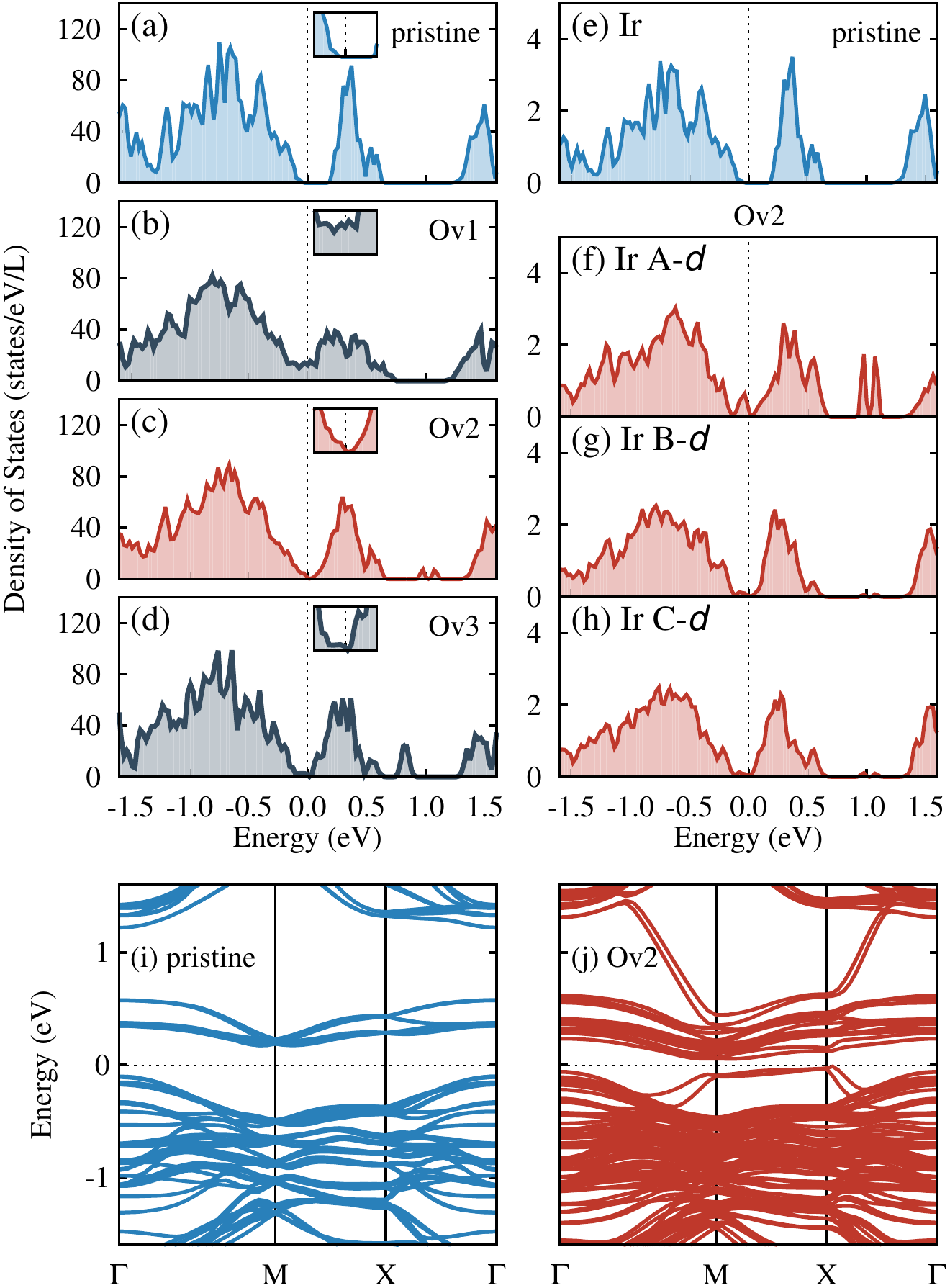}
\caption{(Color online)
(a) Total DOS of  pristine Sr$_3$Ir$_2$O$_7$ structure.
(b)-(d) Total DOS of Sr$_3$Ir$_2$O$_7$ with an oxygen vacancy at position 1 (Ov1), 2 (Ov2), and 3 (Ov3). Insets: magnification of total DOS around the Fermi level.
(e) Ir-$d$ projected DOS for the pristine case. (f)-(h) Ir-$d$ projected DOS of Ir sites A, B, and C for the energetically more favorable Ov2 case (see Fig.~\ref{structure} (b) and (c)).
Bandstructure for (i) pristine case and (j) Ov2 case.
}
\label{dos_ov}
\end{center}
\end{figure}

\begin{figure}[t]
\begin{center}
\includegraphics[width=85mm]{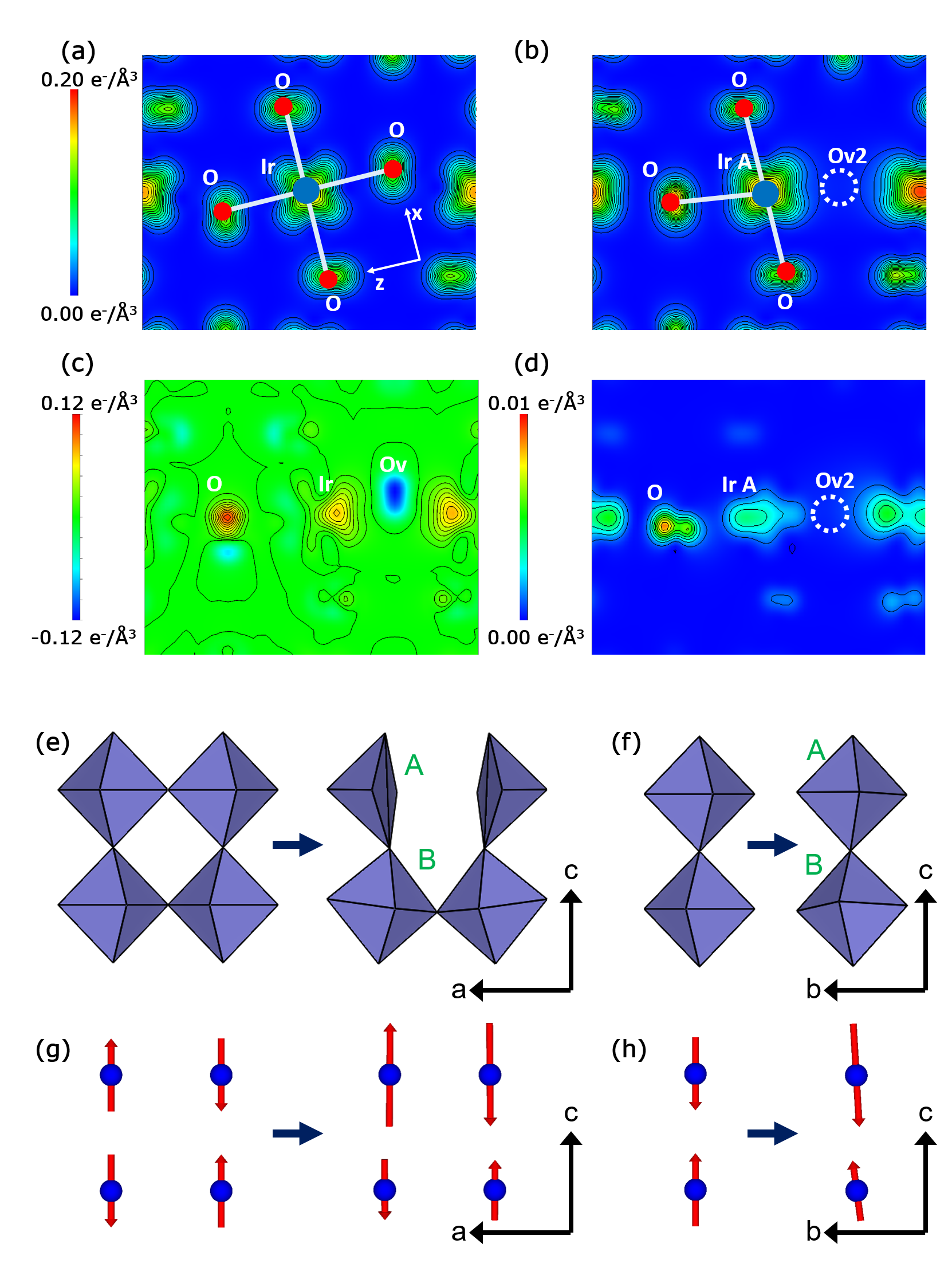}
\caption{(Color online)
Charge density plot of IP layer for the (a) pristine case and (b) Ov2 case which contains Ov site. Energy range is from -2.0 to 0.0 eV with respect to the Fermi level. (a) The octahedron formed from IrO$_6$ is shown for the pristine case with local axis ($x$ and $z$). (b) The square pyramid formed from IrO$_5$ at Ir-A site is shown. (c) Charge density difference between (a) and (b). (d) The charge density plot for the in-gap states just below the Fermi level (-0.12 to 0.0 eV).
(e)-(h) The structural change due to the oxygen vacancies. Side view of the octahedral network perpendicular the (e) $b$- and (f) $a$-direction,
where pristine and Ov2 cases are compared. IrO$_6$ is shown with octahedron. Clear octahedral tilting can be seen after introduction of Ov2.
The local magnetic moment is canted off from the $c$-axis (OP direction) as seen from (g) $b$- and (h) $a$-axis.
The size and direction of each magnetic moment is denoted with the length and direction of the red arrow.
}
\label{cdp}
\end{center}
\end{figure}

The formation of oxygen vacancies affects significantly the structural, electronic and magnetic properties of the system, as elaborated below.
The removal of oxygen atoms in a perovskite destroys the local IrO$_6$ octahedron network, forming square pyramidal shaped IrO$_5$ blocks
as schematically shown in Fig.~\ref{cdp} (a), (b) and (e). Within a simple ionic model, Ovs act as effective electron donors, with each Ov donating formally
two electrons to the system. These excess electrons can be either spread uniformly in the lattice behaving like delocalized charges in the bottom of the conduction band
or can be trapped in specific sites and form defect states in the gap~\cite{Hao2015}. In the delocalized state,
electron doping shifts upwards the chemical potential thereby inducing an insulator-to-metal transition, whereas in the localized solution,
the system remain insulating but shows characteristic localized levels within the gap.

To study the effect of oxygen vacancies on the electronic structure of Sr$_3$Ir$_2$O$_7$ we have calculated the density of states (DOS) for each
Ov type using OP magnetically ordered supercells. In the pristine case, Fig.~\ref{dos_ov} (a), the system is clearly insulating.
Depending on the specific site where the Ov is created the electronic structure exhibits different characteristics: Ov1 corresponds to the delocalized solution
in which the excess electrons fill the conduction band and lead to the closure of the gap thus establishing a metallic state, as shown in Fig.~\ref{dos_ov} (b).
In contrast, within the Ov2 configurations the insulating state is preserved even though the band gap is very small, about 60 meV, see
Fig.~\ref{dos_ov} (c). This state is characterized by a mid-gap peak associated with electrons localized in the Ir-A site as seen from the DOS shown in Fig.~\ref{dos_ov} (f) and from the bandstructure reported in Fig.~\ref{dos_ov} (i) and (j).
Finally, the Ov3 case, represents an intermediate solution between metallic Ov1 and insulating Ov2: the system is formally metallic  but the
density of state at the Fermi level is very small and a band gap could be eventually opened upon charge trapping in some Ir site (Fig.~\ref{dos_ov} (d)).

To clarify the origin and the features of the Ov2 mid-gap peak  we have inspected the charge redistribution induced by  Ov2 in terms
of charge density plots (CDP) projected in the Ir-Ov2 basal $ab$-plane, as reported in Fig.~\ref{cdp} (a)-(d).
The undoped case, Fig.~\ref{cdp} (a), exhibits a charge pattern typical of the Ir$^{4+}$O$_6$ octahedral symmetry: using the local axis indicated inside the picture,
the  $d_{xz}$ orbital at the Ir site is easily recognizable. The formation of Ov2 destroys the local octahedral environment and leads to an accumulation
of the excess electrons at Ir-A and in the surrounding O sites, as clearly seen from the CDP difference between the pristine and Ov2 case shown in
Fig.~\ref{cdp} (c). As a consequence of this charge trapping, the nominally $4+$ oxidation state of Ir-A is reduced to $3+$ and due to the enhanced
Coulomb repulsion between Ir-A and the surrounding oxygens, the Ir-O bond-length is increased by about 5\% from 2.02~\AA{} to 2.13~\AA.
Electron trapping on a transition metal (TM) atom associated with the elongation of the TM-O distance is the typical hallmark of a defect state and explains the
formation of the mid-gap peak in the Ov2 case. This kind of Ov-induced defect state is generally associated either with small polarons, if the trapped charge is
localized within one lattice constant around the trapped center and is loosely bound to the Ov, or, like in this case, forms a Ov-excess electron complex
in which the excess electrons couple with the Ov site.
We now inspect the local orbital character of the defect state.
One would naively expect that the excess electrons would occupy the $t_{2g}$ empty levels at the bottom of the conduction band of the undoped
and undistorted sample (filling the $t_{2g}$ hole of the $J_{eff}$=1/2 state). However,  the disruption of the octahedral environment induces a splitting and re-ordering
of the empty $d$ manifold substantiated by the lowering of the $d_{z^2}$ orbital, which become therefore the lowest unoccupied state available for accommodating the
excess electrons. This is confirmed by the CDP reported in Fig.~\ref{cdp} (d) which shows the characteristic $d_{z^2}$ cigar-like charge lobe along the local $z$ axis.
This charge redistribution process ultimately leads to an increase of the local spin moment at the Ir-A site, from 0.29~$\mu_B$ to 0.51$~\mu_B$, see Table~\ref{moment}.


\begin{table}[b]
\centering
\caption[]{
Local magnetic moment information of Ir ion for pristine and Ov2 case. The unit of the moments are all in $\mu_B$. The IP and OP local magnetic moment ($\mu_{IP}$ and $\mu_{OP}$) denotes total moment. Each spin and orbital moment information ($\mu_{S}$ and $\mu_{O}$) is also provided, with their ratios. For the Ir types, see Fig.\ref{structure}. Ir D denotes the atom from bilayer without Ov (There are two bilayers in a unit cell).
}
\begin{tabular}{C{1.8cm}|C{1.2cm}|C{1.2cm}C{1.2cm}C{1.2cm}C{1.2cm}}
\hline\hline
\multirow{2}{*}{}& \multirow{2}{*}{pristine} & \multicolumn{4}{c}{Ov2}      \\
                        &          & Ir A & Ir B & Ir C & Ir D  \\
\hline
             $\mu_{IP}$        &  0.00    & 0.11 & 0.18 & 0.08 & 0.01    \\
             $\mu_{OP}$        &  0.65    & 0.85 & 0.51 & 0.57 & 0.60    \\
\hline
             $\mu_S$                    &  0.29     & 0.51 & 0.23  & 0.25  & 0.26    \\
             $\mu_L$                    &  0.36     & 0.41 & 0.31  & 0.33  & 0.34   \\
             $\mu_L$/$\mu_S$      &  1.2      & 0.8  & 1.3   & 1.3  & 1.3    \\
\hline\hline
\end{tabular}
\label{moment}
\end{table}

 We show now that the breaking of the local octahedral symmetry and the re-ordering of the $d$ levels in the defective sample also
 influence the value of the local spin and orbital moment,  $\mu_{L}$ and $\mu_{S}$, respectively. The relative value of
 $\mu_{L}$ and $\mu_{S}$, quantified by the ratio $\mu_{L}/\mu_{S}$, is an important quantity in iridates as it can give insights on the
 degree of $J_{eff}$=1/2-ness of the system~\cite{BJKim2008,Jackeli2009,BKim2016_1,BKim2016_2}. The values of $\mu_{L}$ and $\mu_{S}$
 for the undoped and Ov2 case are collected in Tab.~\ref{moment}.
 For the pristine case, the calculated $\mu_{L}/\mu_{S}$ is about 1.2, which is lower than the
 ideal $J_{eff}$=1/2 value of 2, indicating that Sr$_3$Ir$_2$O$_7$ deviates substantially from the ideal $J_{eff}$=1/2 state. For Ov2 case, $\mu_L$/$\mu_S$ for Ir-A is
 further reduced to 0.8, mostly due to the doubling of the spin moment (Table~\ref{moment}), whereas for the other Ir sites it remains almost unchanged
 (1.3). This confirms that the effect of Ovs is very local, essentially circumscribed at the vicinity of the defect state, in line with the CDP analysis and with the conclusions obtained from the DOS: the formation of Ov2 alters the local electronic structure at the Ir-A
 site only (mid-gap state) but leave almost unaltered the DOS at the other Ir sites (see Fig.\ref{dos_ov} (g) and (h)).
 Finally, we note that the magnetic OP order is robust against different types of Ovs and $U$ values as shown in Fig.~\ref{E_U}, implying that our analysis of
 the effect of strain on the electronic and magnetic properties remains valid even in the presence of Ov.

 Now we can discuss on how the formation of oxygen vacancies can explain the origin of the IP magnetic response observed in Sr$_3$Ir$_2$O$_7$~\cite{Boseggia2012}.
 The most crucial role played by the Ovs is the breaking of the octahedra connectivity that affects strongly the local geometry and the magnetic structure, as elaborated below. For the pristine case, despite the strong octahedral rotation within the plane, there is no octahedral tilting along the $c$-axis; as a result the apical Ir-O-Ir angle is 180$^{\circ}$, and it does not even vary upon epitaxial strains as discussed before (see Table~\ref{bond}). When Ov2 vacancies are formed in the planar network of IrO$_6$ bilayers, in addition to the breaking of the octahedral symmetry at the Ir A site, there occurs a substantial apical tilting ($\sim$ 11$^{\circ}$) of the polyhedrons which produce canting-off of the local moment at the Ir-sites as shown in Fig.~\ref{cdp}(e)-(h). Ov2 modifies the overall connectivity of the bilayers and the tilting pattern affects the Ir B and Ir C sites too. As a consequence of this structural rearrangement, the local magnetic moment is canted off from the $c$-axis (OP direction) and a finite IP component of the magnetic moment at the A and B Ir sites emerges, schematically depicted in Fig.~\ref{cdp} (g) and (h). The values of the IP and OP magnetic moments are listed in Tab.~\ref{moment}. The canted moment is slightly larger for Ir B (0.18 $\mu_B$) than Ir A (0.11 $\mu_B$) due to the enhanced tilting of the IrO$_6$ octahedron for the Ir B site. This suggests that the formation of Ovs is indeed responsible for the experimentally observed IP magnetic response of the system, which is unexpected from the global magnetic structure of Sr$_3$Ir$_2$O$_7$.

\section{Conclusions}

 In summary, by means of first principles calculations we have studied two possible mechanisms to modify the magnetic structure of
 the RP $m=2$ iridate Sr$_3$Ir$_2$O$_7$: epitaxial strain and oxygen vacancies. Compressive strain is found to induce a spin-flop phase transition from collinear out-of-plane to canted in-plane magnetic order. The driving force for this unusual transition is the strong dependence of the interlayer exchange interactions on the substrate strain. Compressive strains larger than 3~\% induce a crossover between  the intralayer $J_1$ and interlayer $J_2$ exchange magnetic interactions, mostly attributable  to a huge reduction of the strength of $J_2$ by about one order of magnitude. This result confirms that epitaxial strain is a viable and effective route to tune specific properties of materials, in particular in materials like iridates in which there is a delicate balance between the lattice, spin, and orbital degree of freedom.
 By inspecting the role of oxygen vacancies, we found that the previously reported noteworthy IP magnetic response of Sr$_3$Ir$_2$O$_7$
 should be ascribed to the canting of the local magnetic moment induced by the formation of oxygen vacancies, that perturb the local octahedral network thereby forming defect states and allowing spin flexibility. We hope the further experiments on Sr$_3$Ir$_2$O$_7$ sample with various stoichiometry could confirm our investigations. Moreover, as external stimulus such as strain or intrinsic defect like oxygen vacancies can lead to significant changes of the magnetic order in Sr$_3$Ir$_2$O$_7$, we expect that other types of perturbation such as doping could also induce important modifications on the electronic and magnetic structure of the system.

\begin{acknowledgments}
B.K. thank N.H. Sung for fruitful discussions.
This work was supported by the joint Austrian Science Fund (FWF) and Indian Department of Science and Technology (DST) project INDOX (I1490-N19).
P.L. is grateful to the China Scholarship Council (CSC)-FWF Scholarship Program.
Computing time at the Vienna Scientific Cluster is greatly acknowledged.
\end{acknowledgments}

\bibliographystyle{apsrev4-1}
\bibliography{bibfile}

\end{document}